\documentclass[prb,twocolumn,showpacs,amsmath,amssymb,superscriptaddress]{revtex4-1}

\usepackage{graphicx}%
\usepackage{color} %
\usepackage{bm} %
\usepackage{multirow}%
\usepackage{hyperref} %
\usepackage{bbold} %
\hypersetup{backref=true,
 pdfnewwindow=true, colorlinks=true,
 linkcolor=blue, anchorcolor=blue,
 citecolor=blue, filecolor=blue,
 menucolor=blue, urlcolor=blue}

\begin{document}

\title{Chern-Simons orbital magnetoelectric coupling in generic
  insulators}

\author{Sinisa Coh} 
\email{sinisa@physics.rutgers.edu} 
\author{David Vanderbilt} 
\affiliation{
  Department of Physics and Astronomy, Rutgers University, Piscataway,
  New Jersey 08854-8019, USA
}
\author{Andrei Malashevich}
\author{Ivo Souza}
\affiliation{
Department of Physics, University of California, Berkeley, California 94720,
USA
}

\date{\today}

\def\nn{\nonumber\\}
\def\beq{\begin{equation}}
\def\eeq{\end{equation}}
\def\t{$\theta$}
\def\z2{$\mathbb{Z}_2$}
\def\cro{Cr$_2$O$_3$}
\def\bise{Bi$_2$Se$_3$}
\def\bfo{BiFeO$_3$}
\def\gdo{GdAlO$_3$}
\def\e{{\mathcal E}}
\def\eb{{\bm{\mathcal E}}}
\def\k{{\bf k}}
\def\rr{{\bf r}}
\def\R{{\bf R}}
\def\P{{\bf P}}
\def\O{{\bf 0}}
\def\ti#1{\tilde{#1}}
\def\Ac{{\cal A}}
\def\ket#1{\vert#1\rangle}
\def\bra#1{\langle#1\vert}
\def\ip#1#2{\langle#1\vert#2\rangle}
\def\me#1#2#3{\langle#1\vert#2\vert#3\rangle}
\def\ev#1{\langle#1\rangle}
\def\dint{\displaystyle\int}
\def\bP{\boldsymbol{\mathcal{P}}}
\def\wt#1{\widetilde{#1}}
\def\mone{\mathbf{I}}

\pacs{75.85.+t,03.65.Vf,71.15.Rf}

\begin{abstract}
  We present a Wannier-based method to calculate the Chern-Simons
  orbital magnetoelectric coupling
  in the framework of
  first-principles density-functional theory. In view of recent
  developments in connection with strong \z2\ topological insulators,
  we anticipate that the Chern-Simons contribution to the magnetoelectric
  coupling could, in special cases, be as large or larger than the
  total magnetoelectric coupling in known magnetoelectrics like \cro.
  The results of our calculations for the ordinary
  magnetoelectrics 
  \cro, \bfo\ and \gdo\ confirm that the Chern-Simons contribution is quite
  small in these cases.  On the other hand, we show that if the
  spatial
  inversion and time-reversal symmetries of the \z2\ topological
  insulator \bise\ are broken by hand,
  large induced changes appear in the Chern-Simons magnetoelectric coupling.
\end{abstract}

\maketitle

\section{Introduction}
\label{sec:intro}

In recent years there has been a significant revival of interest
in magnetoelectric effects in solids, as surveyed in several
reviews.\cite{fiebig05,eerenstein06,fiebig09,rivera09}
Potential applications of these materials have long been discussed
\cite{wood75,smolenski82} in areas ranging from the optical
manipulation and frequency conversion to magnetoelectric memories.
Of the various quantities that can be discussed, the linear
magnetoelectric coupling tensor $\alpha_{ij}$ is clearly of
primary interest, as it quantifies the leading-order term in
the coupling at small fields.  
We define it as
\beq 
\label{eq:alpha} 
\alpha_{ij}=\left( \frac{\partial {\cal P}_i}{\partial B_j} \right)_\e
 =\left(\frac{\partial M_j}{\partial \e_i}\right)_B,
\eeq
where ${\cal P}_i$ is the electric polarization induced by the magnetic field
$B_j$, or equivalently, $M_j$ is the magnetization induced by the
electric field $\e_i$.  We use SI units (see Sec.~\ref{sec:units})
and the derivatives are to be evaluated at zero
electric and magnetic field.
In the special case that the induced
response ($\bP$ or $\bf M$) remains parallel to the applied field
($\bf B$ or $\bm{\e}$), the tensor $\bm{\alpha}$ is purely diagonal
with equal diagonal elements, and its strength can be measured by a
dimensionless scalar parameter \t\ defined via
\beq
\label{eq:alpha-iso}
\alpha_{ij}^{\rm iso}=\frac{\theta e^2}{2\pi h}\,\delta_{ij}.
\eeq
More generally, depending on the magnetic point group of the crystal,
$\alpha_{ij}$ can have distinct diagonal components as well as
non-zero off-diagonal ones.

The linear magnetoelectric response $\alpha_{ij}$ can be decomposed into
two contributions coming from purely electronic and from ionic responses
respectively.  The former is defined as the magnetoelectric response
that occurs when atoms are not allowed to displace in response to the
applied field, while the latter is defined as the remaining
lattice-mediated response.  One generally expects ionic effects
to dominate over electronic responses, as for example was shown
recently in Ref.~\onlinecite{iniguez08, delaney09} for the case of \cro.
Moreover, each of these components can be decomposed further into spin
and orbital parts, since the magnetization induced by the electric
field can be decomposed in that way.  Here one would
naively expect that the spin contribution will dominate with respect
to the orbital one, since orbital moments are usually strongly
quenched by crystal fields.  Mostly for this reason,
realistic theoretical calculations of magnetoelectric coupling have
been developed\cite{iniguez08,wojdel09, delaney09} only for the spin
component.

As shown in Refs.~\onlinecite{essin10} and \onlinecite{malashevich10}
using two complementary approaches, the
orbital magnetoelectric polarizability (OMP), defined as the
contribution of orbital currents to the magnetoelectric coupling
$\alpha_{ij}$, can be written as the
sum of three gauge-invariant
contributions. One of these, first discussed by
Qi {\it et al.}\cite{qi08} and Essin {\it et al.},\cite{essin09}
is the Chern-Simons term (CSOMP).  Since this contribution
is purely isotropic, we can measure its strength using
the single parameter \t\ as in Eq.~(\ref{eq:alpha-iso}). In this paper
we will focus mostly on the CSOMP component of $\alpha_{ij}$.  From an
implementation viewpoint, the CSOMP component is quite different from the
other two components of the OMP: it can be calculated from
a knowledge of the ground-state electron wavefunctions alone, but
only after careful attention is given to the need to choose a
smooth gauge in discretized $k$-space.

One of the motivations for the current
work is the possibility of finding a material
whose CSOMP component of the linear magnetoelectric tensor will be
large compared to the total coupling in known magnetoelectric
materials. As elaborated in more detail in Sec.~\ref{sec:back_motiv},
the basis for this possibility arises from the
before-mentioned theoretical developments\cite{fu07-2} and the
experimental verification of the existence of \z2\ topological insulators
such as Bi$_{1-x}$Sb$_x$, Bi$_2$Se$_3$, Bi$_2$Te$_3$ and Sb$_2$Te$_3$.
\cite{hsieh08,hsieh09,xia09}  Roughly speaking, we seek a material
that is similar to a \z2\ topological insulator, but having broken
inversion and time-reversal symmetries. In order to take the first
steps toward searching for such materials, we have set out to
calculate the CSOMP component of the magnetoelectric tensor in
several compounds of interest using density-functional theory.

The paper is organized as follows.
In Sec.~\ref{sec:back_motiv} we provide theoretical background
by reviewing the previously-derived\cite{essin10,malashevich10}
expression for the $\bm{\alpha}$ tensor, and by discussing the
connection between bulk and surface properties in a way that is
analogous to the theory of surface charge and bulk electric polarization.
We also review the connection to \z2\ topological insulators and
make some general comments about symmetry.
In Sec.~\ref{sec:methods} we discuss the gauge-fixing issues that
arise when discretizing the CSOMP expression on a $k$-point mesh,
and show how these can be resolved using Wannier-based methods.
By this route, we arrive at an explicit expression for the
CSOMP in terms of position matrix elements between Wannier functions.
We evaluate this expression in the density-functional context for
several materials of interest in Sec.~\ref{sec:results}.
Finally, we summarize and give an outlook in Sec.~\ref{sec:summary}.

\section{Background and motivation}
\label{sec:back_motiv}

In this section we briefly summarize previous work from
Refs.~\onlinecite{essin10} and \onlinecite{malashevich10} on the orbital
magnetoelectric coupling (OMP), describe relationships between
bulk and surface properties, discuss motivations for this
work based on the discovery of strong \z2\ topological insulators,
and present a brief symmetry analysis.

\subsection{Units and conventions}
\label{sec:units}

In this paper we use SI units and define $\alpha$ according to
Eq.~(\ref{eq:alpha}) using independent field variables $\e$
and $B$.  It follows that $\alpha$ has the same units as the
vacuum admittance $1/c\mu_0$.\cite{hehl09}
While this is convenient from the point of view of first-principles
theory, where $B$ is fixed to zero in practice, the more
conventional definition in the literature is in terms of fixed $\e$
and $H$ fields, in which case one has
\beq
\alpha^{\rm EH}_{ij}=\left( \frac{\partial {\cal P}_i}{\partial H_j} \right)_\e
 =\mu_0\left(\frac{\partial M_j}{\partial \e_i}\right)_H
\label{eq:alpha-eh}
\eeq
and $\alpha^{\rm EH}$ has units of inverse velocity.\cite{rivera94}
In the
typical case that the magnetic susceptibility of the material is
negligible, these are related by $\alpha^{\rm EH}=\alpha\mu_0$,
and one can define a reduced (dimensionless) quantity
$\alpha_{\rm r}=c \mu_0 \alpha=c\alpha^{\rm EH}$.\cite{hehl09}
Defined in
this way, $\alpha_{\rm r}$ is numerically equal to the value
of the magnetoelectric coupling in Gaussian units using the
conventions of Rivera,\cite{rivera94} which in turn corresponds
to the notation ``g.u.'' (``Gaussian units'') in some recent
papers.\cite{iniguez08,wojdel09} Furthermore, using the notation
of Eq.~(\ref{eq:alpha-iso}) 
for the isotropic magnetoelectric coupling,
it follows that 
the diagonal component of
$\alpha_{\rm r}$ is just
$\theta/\pi$ times the fine structure constant (which is 
$e^2 c \mu_0/2h$ in SI units).

\subsection{Theory of orbital magnetoelectric coupling}
\label{sec:omp}

The purely electronic orbital magnetoelectric coupling $\alpha_{ij}$
can be written in terms of three gauge-invariant contributions
\beq
\alpha_{ij}=
\alpha^{\rm CS}_{ij}
+
\widetilde{\alpha}^{\rm LC}_{ij}
+
\widetilde{\alpha}^{\rm IC}_{ij},
\label{eq:alpha-decomp}
\eeq
where $\alpha^{\rm CS}_{ij}=\delta_{ij}\alpha^{\rm CS}$ is the
above-mentioned (isotropic) CSOMP, while
$\widetilde{\alpha}^{\rm LC}_{ij}$ and 
$\widetilde{\alpha}^{\rm IC}_{ij}$ are two additional contributions.
The isotropic 
part of the OMP tensor has contributions from
the two $\widetilde{\alpha}$ terms as well as from the CSOMP term.
The three contributions to the OMP can compactly be expressed as
\begin{align}
&\alpha^{\rm CS}=
\eta \frac{e}{2}
\int
d^3k\, \epsilon_{ijk}\, \mathrm{tr}\left[\Ac_i \partial_j \Ac_k- 
\frac{2i}{3} \Ac_i \Ac_j \Ac_k\right],
\label{eq:alpha-cs}
\\
&\widetilde{\alpha}^{\rm LC}_{ij}=
\eta
\epsilon_{jkl}
\mathrm{Im}
\int d^3k\,
\bra{\wt{\partial}_k u_{n\k}}
(\partial_l H_\k)\ket{\wt{D}_i u_{n\k}},
\label{eq:alpha-lc}
\\
&\widetilde{\alpha}^{\rm IC}_{ij}=
\eta
\epsilon_{jkl}
\mathrm{Im}
\int d^3k\,
  \bra{\wt{\partial}_k u_{n\k}}\wt{D}_i u_{m\k}\rangle 
  \bra{u_{m\k}}(\partial_l H_\k)\ket{u_{n\k}},
\label{eq:alpha-ic}
\end{align}
where the notations are defined as follows. An implied sum notation
applies to repeated Cartesian ($ijkl$) and band ($mn$) indices,
corresponding to a trace over occupied bands in the latter case
(written explicitly as `tr').
A common prefactor $\eta=-e/\hbar(2\pi)^3$ appears in each
equation, with $e>0$ being the magnitude of the electron charge.
The Berry connection
\beq
\Ac_{m n \k j} = \me{u_{m \k}}{i \partial_j }{ u_{n\k}}
\label{eq:berry_conn}
\eeq
is defined in terms of the cell-periodic Bloch functions
\beq
\ket{u_{n \k}} = e^{-i \k\cdot \rr} \ket{\psi_{n \k}},
\eeq
which are the eigenvectors of $H_\k=e^{-i\k\cdot\rr} {\mathcal H}
e^{i\k\cdot\rr}$, where $\mathcal H$ is the
bulk periodic Hamiltonian of the crystal at zero electric and
magnetic field.
$\partial_j$ and $D_j$ are the partial derivatives with respect to
the $j$-th component of the wavevector $\bf k$ and the electric
field $\bm{\e}$ respectively.  Finally, the tilde indicates a
covariant derivative, $\wt\partial_j=Q_\k\partial_j$ and
$\wt D_j=Q_\k D_j$, where $Q_\k= 1-\ket{u_{n\k}}\bra{u_{n\k}}$
(sum implied over $n$).
Additional screening contributions to $\widetilde{\alpha}^{\rm LC}_{ij}$ and 
$\widetilde{\alpha}^{\rm IC}_{ij}$ that occur in the context of
self-consistent field calculations, not given here,
can be found in Ref.~\onlinecite{malashevich10}.

As in the case of electronic polarization, one needs to be careful
about relating the above bulk expressions to experimentally
measurable physical quantities, since arbitrary
surface modifications can contribute to the effective measurable OMP.
The relationship between the OMP and experimentally measurable responses
are explained in more detail in the next section.

\subsection{Relation between bulk and surface properties}

In order to discuss the relationship between bulk and surface
quantities in connection with the OMP, it is instructive first
to review the corresponding connections in the theory
of electric polarization.

\subsubsection{Electric polarization and surface charge}
\label{sec:analogy_P}

We first review the relationship between the bulk electric polarization,
as obtained from the crystal bandstructure according to the
Berry-phase theory,\cite{king-smith,vanderbilt93} and a measurable
quantity which is the macroscopic
dipole moment of a finite sample cut from this crystal.
Given the set of valence
Bloch wavefunctions $\ket{\psi_{n\k}}$ of an insulating
crystal, one can readily calculate the electronic contribution to
the polarization as the integral
\beq 
{\cal P}_i = -\frac{e}{(2\pi)^3} \displaystyle\sum_n \int d^3 k \
   \me{u_{n \k}}{i \partial_{k_i}}{u_{n \k}}
\label{eq:berry_pol}
\eeq
over the Brillouin zone (BZ).
Gauge changes ($\ket{u_{n \k}}\rightarrow
e^{-i\beta(\k)}\ket{u_{n \k}}$) 
can change the value of this
integral only by ${\bf R} e/ \Omega$,
where $\bf R$ is a lattice vector and $\Omega$ is the unit cell volume.
The value of this integral is therefore only well-defined modulo
${\bf R} e/ \Omega$.  In what follows we assume that a definite choice
of gauge has been made so that a definite value of $\bP$ has been
established.  We now analyze how, and under what circumstances, one
can relate this $\bP$ to the (experimentally measurable) dipole
moment $\bf d$ of an arbitrarily faceted finite sample of this crystal.

At each local region on the surface of this finite sample, assuming
a perfect surface preparation (defect-free with ideal periodicity),
we can relate $\bP$ to the surface charge density $\sigma$ at that same
point via \cite{vanderbilt93}
\beq 
\sigma = \left( \bP + \frac{e}{\Omega}{\bf R}
\right) \cdot
\hat{n} + \Delta .
\label{eq:surf_thm}
\eeq
Here $\hat{n}$ is the surface normal unit vector, $\bf R$ is a lattice
vector, and $\Delta$ is an additional contribution present only
for metallic surfaces.
The term involving $\bf R$, which corresponds to an integer number
of electrons per surface unit cell, is required because, for a
given surface $\hat{n}$, it may be possible to prepare the surface in
different ways (e.g., by adding or subtracting a layer of ions,
or by filling or emptying a surface band) such that the surface
charge per cell changes by a quantum.  Thus, $\bf R$ is in general
a surface-dependent quantity in Eq.~(\ref{eq:surf_thm}).
If the surface patch under consideration is not insulating,
then $\Delta$ is a term which measures the contribution of
the partially occupied surface bands to the surface charge, and is
proportional to the area fraction of occupied band in $k$ space.
(In the case of an insulator with non-zero first Chern number, this
fraction has to be calculated with special care,\cite{coh09}
but we shall not consider this case in what follows.)

Now, let us consider the special case that all surfaces are
insulating ($\Delta=0$) and that the surface charges of {\it all}
surface patches are consistent with a {\it single} vector value of
$\bf R$ (``global consistency'').  Under these circumstances, the
macroscopic dipole moment $\bf d$ of the crystallite is given by
\beq
{\bf d} = {\cal V} \left(\bP + \frac{e}{\Omega} {\bf R} \right),
\label{eq:deq}
\eeq
which can be obtained trivially by integrating Eq.~(\ref{eq:surf_thm}).
Here $\cal V$ is the volume of entire finite sample.
As could be anticipated, ${\bf d}/{\cal V}$ has a component
depending only on the bulk wavefunctions and our gauge choice,
and an additional component $e{\bf R}/\Omega$ reflecting the
preparation of the surfaces.

\subsubsection{OMP and surface anomalous Hall conductivity}
\label{sec:analogy_OMP}

We now discuss a corresponding set of relationships between the
bulk-calculated OMP and the surface anomalous Hall conductivity.

Using Eqs.~(\ref{eq:alpha-cs}), (\ref{eq:alpha-lc}), and
(\ref{eq:alpha-ic})
one can calculate the tensor ${\bm \alpha}$ from the knowledge of bulk
Hamiltonian of an insulating crystal. Analogously as in the case of
polarization, one can again show that a gauge change
\footnote{Here we refer to a multiband gauge transformation
  having the form of Eq.~(\ref{eq:gauge}), since $\bm\alpha$ can
  be shown to be fully invariant under a single-band phase
  twist.}
must either
leave  $\bm \alpha$ invariant or change it by a quantum
$m(e^2/h)\mone$,
where $m$ is an integer and $\mone$ is the unit matrix.
More precisely,
this gauge transformation will only affect the CSOMP component
${\bm \alpha}^{\rm CS}$ of the OMP, since the other two contributions
$\widetilde{\bm \alpha}^{\rm LC}$ and $\widetilde{\bm \alpha}^{\rm IC}$
are fully gauge-invariant (see Ref.~\onlinecite{malashevich10} for details).

We now imagine cutting a finite crystallite from this infinite crystal,
and we wish to relate ${\bm \alpha}$ to its physically observable
linear magnetoelectric coupling ${\bm \beta}$, defined for a
finite sample by
\beq 
\beta_{ij}=\frac{\partial d_i}{\partial B_j} 
          =\frac{\partial \mu_j}{\partial \e_i},
\eeq
where $d_i$ is the dipole moment of the finite sample and $\mu_j$ is
its magnetic dipole moment. We want to discuss this relationship in a
way that is analogous to that between the bulk $\bP$  and sample
dipole moment $\bf d$ in Sec.~\ref{sec:analogy_P}.

As follows from Eq.~(\ref{eq:alpha}),
the application of
an electric field $\e_j$ to the insulating crystal induces the
magnetization
\beq
M_k = \alpha_{jk} \e_j,
\eeq
where $\bm \alpha$ is given by Eq.~(\ref{eq:alpha-decomp}) and is
only determined modulo the quantum $m(e^2/h)\mone$.
Having a
homogeneous $M_k$ inside
the sample and $M_k=0$ outside is equivalent to having a
surface current $K_i$ equal to
\beq
K_i = \epsilon_{ikl} M_k n_l,
\eeq
where $n_l$ is the surface unit normal.
By eliminating $M_k$ from these equations, we see that having
a magnetoelectric tensor $\bm \alpha$ is equivalent to having a surface
anomalous Hall conductivity
$\sigma^{\rm AH}_{ij} = \epsilon_{ikl} \alpha_{jk} n_l$.
If the surface patch in question is insulating, then its
anomalous Hall conductivity should just be given,
modulo
$m(e^2/h)\mone$, by this equation.  If instead the surface
patch is metallic, then an additional surface contribution
$\Delta_{ij}$ should be present, leading to the relation
\beq
\sigma^{\rm AH}_{ij} = \epsilon_{ikl} \left( \alpha_{jk} + m \frac{e^2}{h} 
\delta_{jk} \right) n_l + \Delta_{ij}.
\label{eq:surf_cond}
\eeq
This equation is in precise analogy to Eq.~(\ref{eq:surf_thm})
relating the polarization to the surface charge.
Here $\Delta_{ij}$ may in general contain dissipative contributions,
but in the dirty limit
it will be dominated by the
intrinsic surface contribution 
that can be calculated as a 2D BZ
integral of the Berry curvature of the occupied surface
states.\cite{haldane04}
The integer quantum $m$ appearing in Eq.~(\ref{eq:surf_cond})
corresponds to the theoretical possibility that the surface preparation can
be changed in such a way that a surface band having a nonzero
Chern number may become occupied.
For example, this could be done in principle by constructing a 2D
quantum anomalous Hall layer (as described, e.g., by the Haldane
model\cite{haldane88}), straining it to be commensurate with the
surface, and adiabatically turning on hopping matrix elements to
``stitch it'' onto the surface.

In the special case that all surface patches are insulating ($\Delta_{ij}=0$),
and all surface patches have an anomalous Hall conductivity given by
Eq.~(\ref{eq:surf_cond}) with the {\it same} value of $m$
(``global consistency''), we can relate the
experimentally measurable magnetoelectric response $\bm \beta$ of the
finite crystallite to the bulk-calculated $\bm \alpha$ via
\beq
{\bm \beta} = {\cal V} \left( {\bm \alpha} + m \frac{e^2}{h} \mone \right),
\label{eq:beta_alpha}
\eeq
which follows by integrating Eq.~(\ref{eq:surf_cond}) over all
surfaces.  This equation is in close analogy to Eq.~(\ref{eq:deq})
for the case of electric polarization.  In particular, we see that
${\bm \beta}/{\cal V}$ has a component $\bm\alpha$ depending only
on the bulk wavefunctions and our gauge choice, and an additional
component that is an integer multiple of $(e^2/h)\mone$,
reflecting the preparation of the surfaces.

As will be discussed in the next section, time-reversal symmetry
imposes additional constraints on $\bm \alpha$, and some care
is needed in the interpretation of Eq.~(\ref{eq:beta_alpha})
for the case of \z2\ topological insulators.

\subsection{Motivation and relationship to strong \z2\ topological insulators}
\label{sec:z2}

In this Section, we give arguments to motivate our hope that in
certain materials the CSOMP might be on the order of, or even
much larger than, the total magnetoelectric coupling in typical
known magnetoelectric materials.  For simplicity, we focus henceforth
only on the CSOMP part of the total OMP response, even
though there are additional contributions coming from
$\widetilde{\alpha}^{\rm LC}$ and $\widetilde{\alpha}^{\rm IC}$.
Thus, from now on, the quantity \t\ measures the strength of the
CSOMP through the relation $\alpha^{\rm CS}=\theta e^2/2\pi h$.

\subsubsection{Time-reversal symmetry constraints on \t}

Let us analyze the allowed values of \t\ for an infinite bulk
insulating system that respects time-reversal ($T$) symmetry.  Since
$T$ flips the sign of the magnetic field, it will also reverse the
sign of \t.  As mentioned earlier in Sec.~\ref{sec:analogy_OMP}, 
however, the value of $\theta$ can be changed by $2\pi$ under
a gauge transformation. Therefore one concludes\cite{qi08,essin09}
that the allowed
values of \t\ consistent with $T$ symmetry are $0$ mod $2\pi$
and $\pi$ mod $2\pi$, and that these two cases provide a topological
classification of all $T$-invariant insulators.
Indeed, this classification has been shown\cite{qi08,essin09} to be
identical to the one based on the \z2\ index, with
\z2-odd or ``strong topological'' insulators having $\theta=\pi$,
while \z2-even or ``normal'' insulators have $\theta=0$, even though
the \z2\ index is most often introduced in a different context.\cite{hasan10}
(Incidentally,
$\widetilde{\bm \alpha}^{\rm LC}=\widetilde{\bm \alpha}^{\rm IC}=0$
in both cases since these terms are fully gauge-independent,
unlike the CSOMP term which can be changed by $2 \pi$.)

Consider now
a finite sample of a normal
(\z2-even) $T$-symmetric insulator ($\theta=0$ in the bulk)
with insulating surfaces ($\Delta_{ij}=0$)
prepared in a way that 
the integer 
$m$ is nonzero, and the same on every surface.  From
Eq.~(\ref{eq:beta_alpha}) we conclude that this sample will
have a non-zero magnetoelectric response, $\bm \beta$, proportional to
$m$. Obviously a sample that has $T$ symmetry both in the bulk and on
the surface must have ${\bm \beta}=0$, and therefore we conclude
that this system needs to have broken $T$ reversal symmetry at the
surface. As mentioned earlier,
one could, at least formally, prepare such a surface by
starting from the one that has $m=0$ and then absorbing to each
surface a layer of anomalous Hall insulator\cite{haldane88} with
Chern index $m$. Such a procedure will keep the surfaces insulating
but it will necessarily break the $T$-reversal symmetry.

\begin{figure}
\centering\includegraphics{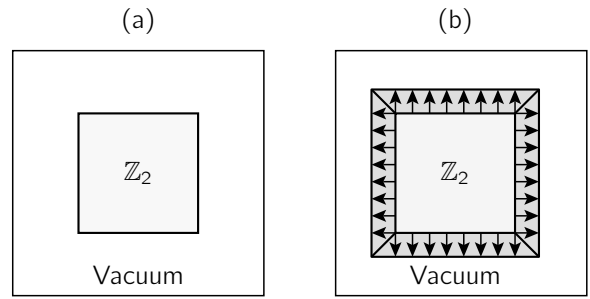}
\caption{Identical samples cut from a strong \z2\ topological insulator,
  but with two different surface preparations.  (a) Time-reversal
  symmetry is preserved at vacuum-terminated surfaces; the
  net magnetoelectric coupling of this sample is zero. (b)
  Time-reversal symmetry is broken at the surface as a result
  of exchange coupling to an insulating ferromagnetic adlayer;
  if this opens a gap in the surface-state spectrum, the entire
  sample will behave as if it has a magnetoelectric coupling of
  exactly $\theta=\pi$.}
\label{fig:fm}
\end{figure}

Next we
analyze the case of a strong \z2\ topological
insulator having $\theta=\pi$, or equivalently, 
${\bm \alpha}={\bm \alpha}^{\rm CS}=(e^2/2h)\,\mone$.
We first consider a sample of such a system that has $T$ symmetry
conserved at its surfaces, as in Fig.~\ref{fig:fm}(a). Again, since the entire
sample is $T$-symmetric, its experimentally measurable magnetoelectric
coupling tensor $\bm \beta$ clearly has to vanish.  Using
Eq.~(\ref{eq:surf_cond}) and the fact that $m$ can take on only
integer, and not half-integer, values, we conclude that the only
way to make the response of the entire sample vanish is to have
$\Delta_{ij}$ be non-zero.  This requires that the surfaces of
such a system must be metallic.  Moreover, since the contribution
$\Delta_{ij}$ of the metallic surface band to the surface anomalous
Hall conductivity is just given by the Berry phase around the
Fermi loop,\cite{haldane04} the needed cancellation requires
this Berry phase to be exactly $\pm \pi$. All this is in precise
accord with the known properties of \z2-odd insulators
and their topologically protected surface states.\cite{hasan10}

The Kramers degeneracy at the Dirac cone in the surface bandstructure
can be removed by the application of a $T$-breaking perturbation to
the surface. In principle, this could be accomplished, for example, by
applying a local magnetic field to the surface or by interfacing the
surface to an insulating magnetic overlayer.  In the latter case,
the interatomic exchange couplings provide a kind of effective
magnetic field acting on the surface layer of the topological
insulator.  If the local Fermi
level resides in the gap opened by field,
then the surface becomes insulating.  If the field
can be consistently
oriented (see Ref.~\onlinecite{qi08}) on each patch of the surface,
either along or opposite the direction of surface normal vector $\bf
n$ (as shown in Fig.~\ref{fig:fm}(b)), then the entire surface becomes
insulating. It is important that the field is applied consistently in
the same direction with respect to $\bf n$, since conducting channels
will otherwise appear at domain boundaries.\cite{hasan10}

If all of these requirements are met, the surface contribution
$\Delta_{ij}$ to $\bm \beta$ vanishes, so that ${\bm \beta} = {\cal V}
{\bm \alpha}$ with $\bm \alpha$ given only by bulk value of
$\theta=\pi$ (assuming $m=0$ for simplicity). Therefore such a sample
of a strong \z2\ topological insulator would behave as if the entire
sample has exactly half a quantum of magnetoelectric coupling
($\theta=\pi$), even though its bulk is time-reversal symmetric!

\subsubsection{Prospects for large-$\theta$ materials}
\label{sec:largetheta}

Recently surface-sensitive ARPES measurements have experimentally confirmed
that several compounds,\cite{hsieh08,hsieh09,xia09}
including Bi$_{1-x}$Sb$_x$, Bi$_2$Se$_3$, Bi$_2$Te$_3$ and Sb$_2$Te$_3$,
do indeed behave as strong \z2\ topological insulators. Therefore
their bulk wavefunctions must be characterized by $\theta=\pi$. Up to
now, the corresponding magnetoelectric response has not been measured
experimentally, in part because of the difficulties in obtaining
truly insulating behavior in the bulk, as well as the need to gap
the surfaces by putting them in contact with magnetic overlayers as
described earlier.

We believe that a more promising approach to observing a large
CSOMP (i.e., \t\ comparable to $\pi$) is to consider an insulator
that has neither $T$ nor spatial inversion symmetry. In this case
the \z2\ classification does not apply, and the surface can be
gapped without any need to apply a $T$-breaking perturbation.
(A more precise statement of the symmetry considerations will be
given in Sec.~\ref{sec:symmetry}.)  The sample can then display a
bulk magnetoelectric coupling of the simple form
${\bm\beta}={\cal V}{\bm\alpha}$.
We note that an orbital
magnetoelectric coupling of $\theta\simeq\pi$ (i.e.,
$\alpha_{\rm r}\simeq1/137$) would correspond to $\alpha^{\rm
EH}\simeq 24.3$\,ps/m, a value that is significantly larger than
the observed coupling in \cro, one of the best-studied magnetoelectric
materials.  For comparison, the reported experimental values for
$\alpha^{\rm EH}_{\perp}$ in \cro, which are presumably dominated
by spin-lattice coupling, range between 0.7 and 1.6\,ps/m at
4.2\,K.\cite{wiegelmann94,kita79}

Of course, in order to have a good
chance of finding a material with a large \t, it may be advisable to
look for materials with some of the same characteristics as the
known \z2-odd insulators, of which the most important is probably
the presence of heavy atoms with strong spin-orbit coupling.
We see no strong reason why such a search might not reveal
a material having a large OMP in the above sense.

\begin{figure}
\vspace{0.5cm}
\centering\includegraphics{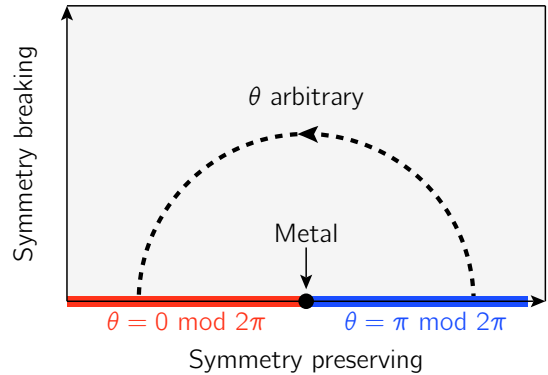}
\caption{(Color online) Schematic view of the allowable values of $\theta$ in
  different parts of the two-parameter space of some unspecified
  model Hamiltonian.  
  Horizontal axis corresponds to the perturbation that preserves at
  least one of the symmetries that render \t\ to be $0$ or $\pi$ (see
  Sec.~\ref{sec:symmetry}). Vertical axis parameterizes a perturbation
  that breaks those symmetries and allows \t\ to be arbitrary.
  See text for the details.}
\label{fig:diag}
\end{figure}
To illustrate the kind of a search we have in mind,
consider some model Hamiltonian that depends on two parameters, one
that preserves either the $T$ or spatial inversion symmetry (or both),
and another that that breaks symmetry such that \t\ takes a
generic value. The possible behavior of such a model is sketched
in Fig.~\ref{fig:diag}, where
these two parameters are plotted along the horizontal
and vertical axes respectively.  The figure also indicates the
generic value of \t\ in each region of parameter space.  Along the
horizontal axis, where the extra symmetry is present, three regions
are indicated. The black dot indicates a point of gap closure
forming the boundary between a normal $T$-symmetric insulator
regime on the left ($\theta=0$) and a strong \z2\ topological
insulator regime on the right ($\theta=\pi$). If the system is
carried along the horizontal axis, $\theta$ must be either $0$ or
$\pi$ except at the critical point, and it must therefore jump
discontinuously when passing through this point of metallic
behavior.  On the other hand, if we now imagine passing from
the \z2-odd to the \z2-even phase along the dashed curve
in Fig.~\ref{fig:diag}, \t\ can vary smoothly and continuously
from $\pi$ to 0 without any gap closure anywhere along the path.
If we can identify a material lying near, but not at, the
right end of this dashed path, it could be the kind of
large-\t\ material we seek.

Thus, our ultimate goal is to use first-principles calculations to
search for a large $\theta$, not in a topological insulator, but
in an ``ordinary'' (but presumably strongly spin-orbit coupled)
insulating magnetic material.  While our work has yet to result
in the identification of a large-\t\ material of this kind,
it represents a first step in the desired direction.

\subsection{General symmetry considerations}
\label{sec:symmetry}

Recall that $\theta$ is a pseudoscalar that changes sign under
time-reversal and spatial-inversion symmetries (since $\bf B$ changes
sign under $T$ while $\bm{\e}$ changes sign under inversion).
On the other hand, %
$\theta$ is invariant under any
translation or proper rotation of a crystal.
Therefore if the magnetic point group of a crystal contains
an element that involves $T$, possibly combined with a proper
rotation, the value of $\theta$ is constrained to be $0$ or $\pi$
(modulo $2\pi$) as discussed earlier. The same happens if the
magnetic point group contains inversion symmetry or any other
improper rotation.

All 32 of the 122 magnetic point groups that do not contain such
symmetry elements, and which therefore allow for an arbitrary value of
$\theta$, are listed in Table~\ref{tab:symm}.
(The bold entries in the table are those magnetic groups for which
the tensor ${\bm \alpha}$ must be isotropic, i.e., a constant times
the identity matrix; the same magnetic groups were also analyzed
in Ref.~\onlinecite{hehl09}).
Clearly we can constrain our
search for interesting materials to the cases listed in the Table.

\begin{table}
  \caption{\label{tab:symm} Magnetic point groups for which a non-zero
    CSOMP is allowed by symmetry.  Notation follows
    Ref.~\onlinecite{cracknell}. 
    Point groups in bold allow only for a purely isotropic
    magnetoelectric tensor.}
\begin{ruledtabular}
\begin{tabular}{lllllll}
$1$ & $\bar{1}'$ & $2$ & $m'$ & $2/m'$ & $222$ & $m'm'2$ \\ 
$m'm'm'$ & $4$ & $\bar{4}'$ & $4/m'$ & $3$ & $\bar{3}'$ & $6$ \\
$\bar{6}'$ & $6/m'$ & $422$ & $4m'm'$ & $\bar{4}'2m'$ & $4/m'm'm'$ & $32$ \\
$3m'$ & $\bar{3}'m'$ & $622$ & $6m'm'$ & $\bar{6}'m'2$ & $6/m'm'm'$ & \\
$\bf{23}$ & $\bf{m'3}$ & $\bf{432}$ & $\bf{\bar{4}'3m'}$ &
$\bf{m'3m'}$ & & \\
\end{tabular}
\end{ruledtabular}
\end{table}

\section{Methods}
\label{sec:methods}

In this section we present our methods for calculating the CSOMP in
the framework of density-functional theory, and analyze in more detail
its mathematical properties and the formal similarities to the formulas
used to calculate electric polarization
and anomalous Hall conductivity.

\subsection{Review of Berry formalism}
\label{sec:geom}

Assume we are given the Bloch wavefunctions
$\ket{\psi_{n \k}}=e^{i \k\cdot \rr}\ket{u_{n\k}}$
as a function of wavevector $\k$ in the $d$-dimensional BZ
($d=1$, 2, or 3) for an insulator having valence bands indexed
by $n\in\{1,\ldots,N\}$.  We work with the cell-periodic Bloch
functions $u_{n \k}(\rr)=e^{-i \k\cdot \rr}\psi_{n\k}(\rr)$
and allow them to be mixed at each
$k$ point by an arbitrary $k$-dependent unitary matrix
\beq
\ket{u_{n \k}} \rightarrow \ket{u_{m \k}} U_{mn\k}
\label{eq:gauge}
\eeq
(sum on $m$ implied).
After this gauge transformation the
wavefunctions are no longer eigenfunctions of the
Hamiltonian, but they span the same $N$-dimensional subset
of the Hilbert space as the true eigenfunctions.
For any given choice of gauge, we define the Berry connection
\beq
\Ac_{m n \k j} = \me{u_{m \k}}{i \frac{\partial}{\partial k_j} }{ u_{n\k}},
\label{eq:connection}
\eeq
which is a $k$-dependent $N \times N \times d$ matrix that measures,
at each $k$ point, the infinitesimal phase difference
between the $m$-th and $n$-th wavefunctions associated with
neighboring points along Cartesian direction $j$ in $k$ space.
This object was already
briefly introduced in Eq.~(\ref{eq:berry_conn}).

In the context of electronic structure calculations, we can now
list three material properties that can be evaluated knowing only
the Berry connection: the electric polarization,
the intrinsic anomalous Hall conductivity, and the CSOMP.

The electric polarization $\cal P$ already appears in dimension
$d=1$ and it can be evaluated as an integral of the Berry
connection over the one-dimensional BZ as \cite{king-smith}
\beq
{\cal P} = -\frac{e}{2\pi} \dint_{\rm BZ} d k \, {\rm tr} \Ac_{k},
\label{eq:pol}
\eeq
where the trace is performed over the band indices of the Berry
connection, as in Eq.~(\ref{eq:berry_pol}).
The integrand is also referred to as the Chern-Simons 1-form,
and its integral over the BZ is well known to be defined
only modulo $2\pi$. Any periodic adiabatic evolution of the Hamiltonian
${\mathcal H}(\lambda)$ whose first Chern number in ($k,\lambda$) space is
non-zero will change the integral above by a multiple of
$2\pi$.\cite{king-smith}

Unlike one-dimensional systems, crystals in $d=2$ can have an
anomalous Hall conductivity.  For a metal, the intrinsic
contribution from a band crossing the Fermi level
can be evaluated as a %
line integral\cite{haldane04,xinjie06}
\beq
 \sigma^{\rm AH} = \frac{e^2}{h} \frac{1}{2\pi}
 \displaystyle\oint_{\rm FL} d \k \cdot {\bm \Ac}_{\k}
\label{eq:ahc}
\eeq
over the Fermi loop.
Fully-filled deeper bands can also make a quantized contribution
given by a similar integral, but around the entire BZ;
this is the only contribution in the case of a quantum anomalous
Hall insulator.\cite{haldane88}
(In both cases, the gauge choice on the boundary of the region should
be consistent with a continuous, but not necessarily $k$-periodic,
gauge in its interior; alternatively, each expression can be
converted to an area integral of a Berry curvature to resolve
any uncertainty about branch choice. See Ref.~\onlinecite{xinjie07}
for more details.)

Finally, unlike one- or two-dimensional systems, three-dimensional
systems can have an isotropic magnetoelectric coupling. The CSOMP
can be evaluated in $d=3$ as a BZ integration
of a quantity involving the Berry connection:
\beq 
\theta = -\frac{1}{4\pi} \dint_{\rm BZ} d^3 k \epsilon_{ijk} {\rm tr} 
\left[ \Ac_i \partial_j \Ac_k -\frac{2i}{3} \Ac_i \Ac_j \Ac_k
\right].
\label{eq:theta}
\eeq
The integrand in this expression is known as the
Chern-Simons 3-form, and its integral over the entire BZ
is again ill-defined modulo $2\pi$, since any periodic adiabatic
evolution of the Hamiltonian ${\mathcal H}(\lambda)$ whose second
Chern number in $(\k,\lambda)$ space is non-zero will change $\theta$
by an integer multiple of $2\pi$.\cite{qi08,essin09}

\begin{figure}
\centering\includegraphics{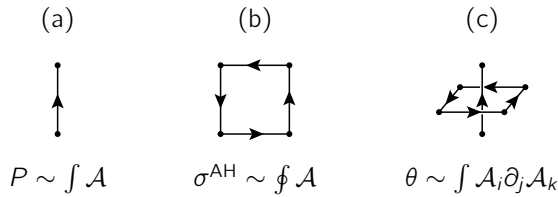}
\caption{Graphical interpretation of Eqs.~(\ref{eq:pol}) (a),
  (\ref{eq:ahc}) (b) and (\ref{eq:theta}) (c) in the case of one occupied
  electron band and for cubic crystal symmetry, for simplicity. See
  text for more detail.}
\label{fig:cs}
\end{figure}

The sketches in Fig.~\ref{fig:cs} compare the geometrical
characters of the operations needed to evaluate
Eqs.~(\ref{eq:pol}-\ref{eq:theta}) in practice.  We consider
the case of one occupied electron band for simplicity.
The polarization of Eq.~(\ref{eq:pol}) is calculated by a line
integral; on a discrete $k$-mesh, the integral of the Berry
connection $\Ac$ over each of line segment, as in
Fig.~\ref{fig:cs}(a), is converted to a discretized form (see
Eq.~(\ref{eq:disc_pol})).
Similarly, in two dimensions the anomalous Hall conductivity of
Eq.~(\ref{eq:ahc}) can be calculated as suggested in Fig.~\ref{fig:cs}(b)
by dividing the occupied part of the BZ into small
square segments and then integrating $\Ac$ around each square.
(Equivalently, one can integrate
$\Ac$ along the Fermi loop.\cite{xinjie07})
In three dimensions,
Fig.~\ref{fig:cs}(c), Eq.~(\ref{eq:theta}) can
be evaluated by dividing the BZ into small cubes.
In each, one needs to multiply the integral of $\Ac$ along
one of the Cartesian directions (as in Eq.~(\ref{eq:pol})) with the
integral of Berry connection in the square orthogonal to that
direction (as in Eq.~(\ref{eq:ahc})), followed by a
symmetrization over the three Cartesian directions.

\subsection{Numerical evaluation of \t }
\label{sec:theta_calc}

In electronic-structure calculations, the cell-periodic wavefunctions
$\ket{u_{n \k}}$ are typically calculated on a uniform $k$-space
grid with no special gauge choice; in general, one should assume
that the phases have been randomly assigned.  Nevertheless, it is
straightforward to construct a gauge-invariant polarization formula
that is immune to this kind of scrambling of the gauge.\cite{mv97}
In one dimension with $k_j$ for $j\in \{1,\ldots,M \}$ (where $k_M$
is the periodic image of point $k_1$), the electronic polarization
is calculated as
\beq
P = \frac{e}{2 \pi} {\rm Im} \ln 
\det \left[
M_{k_1 k_2}  M_{k_2 k_3} ... M_{k_{M-1} k_M}
\right]
\label{eq:disc_pol}
\eeq
where the overlap matrix $M_{k k'}$ is defined as
\beq
\left[ M_{k k'} \right]_{mn} = \ip{u_{m k}}{u_{n k'}}.
\eeq
The reason for using Eq.~(\ref{eq:disc_pol}) is that 
the determinant of the matrix $M_{k_1 k_2}  M_{k_2 k_3} ... M_{k_{M-1} k_M}$ 
is gauge-invariant under any transformation in
the form of Eq.~(\ref{eq:gauge}).
Additionally, the implementation of Eq.~(\ref{eq:disc_pol}) is numerically
stable even when there are band crossings.
A similar gauge-invariant discretization can also be used to calculate
the anomalous Hall conductivity $\sigma^{\rm AH}$.\cite{xinjie07}

Unfortunately, except in the single-band (``Abelian'') case, we are
unaware of any corresponding gauge-invariant discretized formula for
the integral of the Chern-Simons 3-form.  As a result, we have no
prescription for computing the CSOMP that is exactly gauge-invariant
for a given choice of $k$ mesh.  This is a serious problem. Unlike
the calculation of the polarization, which is straightforward even if
the gauge is randomly scrambled at each mesh point, the calculation of
the CSOMP requires that we first identify a reasonably smooth gauge on
the discrete mesh.

The problem of finding a smooth gauge in $\k$ is essentially the same
as that of finding well-localized Wannier functions.  For this
reason, we have adopted here the approach of first
constructing a Wannier representation for the valence bands, and
then using it to compute the CSOMP.  In fact, starting from
Eq.~(\ref{eq:theta}), we derive an expression that allows us to
compute \t\ directly in the Wannier representation.  Once we have
well-localized Wannier functions, this guarantees smoothness of the gauge
and avoids problems with band crossings. Admittedly,
such a formula still depends on the gauge choice, meaning that
different choices of Wannier functions will lead to slightly different
results. However, this difference will vanish as one increases the
density of the $k$-point mesh, since in the continuum limit
the $k$-space expression for \t\ is gauge-invariant (modulo
$2\pi$). More precisely, we expect the calculation of $\theta$ to
converge once the inverse of the $k$-point mesh spacing becomes much
larger than the spread of the Wannier functions.

Therefore, we adopt the strategy of calculating $\theta$ on $k$ meshes of
different density, and extrapolating $\theta$ to the limit of an
infinitely dense mesh. Furthermore, we construct
maximally-localized Wannier functions (MLWF) following Ref.~\onlinecite{mv97},
expecting this to give relatively rapid
convergence as a function of the $k$ mesh density.

Recall that the Wannier function associated with (generalized) band
index $n$ in unit cell $\R$ is defined in terms of
the rotated Bloch states (\ref{eq:gauge}) as
\beq 
\ket{\R n}=\frac{\Omega}{(2\pi)^3}\int d^3 k \,
e^{i\k\cdot(\rr-\R)}\ket{u_{m \k}}U_{mn\k} .
\label{eq:wf}
\eeq 
In the case of MLWFs, the $U_{mn\k}$ are chosen in such a way that the
total quadratic spread of the Wannier function is
minimized.\cite{mv97} 
(In practice the BZ integral is replaced by a summation
over a uniform grid of $k$ points.)

Using Eq.~(\ref{eq:wf}), one can relate the
Berry-connection matrix $\Ac_{mn\k j}$ in the smooth
gauge to the Wannier matrix elements of the position
operator through\cite{mv97}
\beq
\label{eq:A-R}
A_{mn\k j} = \sum_{\R}e^{i\k\cdot\R}\bra{\O m}r_j\ket{\R n}.
\eeq
Replacing each occurrence of $A_j$ in Eq.~(\ref{eq:theta}) with the
above gives, after some algebra,
\begin{align}
\theta
=&
\frac{1}{4 \pi} \frac{(2 \pi)^3}{\Omega} \epsilon_{ijk}
{\rm Im}
\bigg[
\frac{1}{3} \sum_\R \me{\O m}{r_i}{\R n} \me{\R n}{r_j}{\O m} R_k \notag \\
&-
\frac{2}{3} \sum_{\R\P}
     \me{\O l}{r_i}{\R m} \me{\R m}{r_j}{\P n} \me{\P n}{r_k}{\O l}
\bigg]
,
\label{eq:csomp1}
\end{align}
where the sum is implied over band ($lmn$)
and Cartesian ($ijk$) indices.

To obtain a more symmetric form, we introduce a modified
position-operator matrix element between WFs defined as
\beq
\me{\R m}{\ti{r}_i}{\P n}
=\me{\R m}{r_i}{\P n} \left( 1 - \delta_{mn} \delta_{\R \P} \right)
\eeq
and a notation for the Wannier center
\beq
\tau_{ni} = \me{\O n}{r_i}{\O n}.
\eeq
Then Eq.~(\ref{eq:csomp1}) becomes
\begin{align}
\theta
=&
\frac{1}{4 \pi} \frac{(2 \pi)^3}{\Omega} \epsilon_{ijk}
\;\times \\
& \; {\rm Im} \bigg[ \sum_\R
\me{\O m}{\ti{r}_i}{\R n} \me{\R n}{\ti{r}_j}{\O m} 
\left( R_k + \tau_{nk} - \tau_{mk} \right) \notag \\
&\;-
\sum_{\R\P}
\frac{2}{3} \me{\O l}{\ti{r}_i}{\R m} \me{\R m}{\ti{r}_j}{\P n} 
\me{\P n}{\ti{r}_k}{\O l}
\bigg]
.
\label{eq:csomp2}
\end{align}
We find this form more convenient because it separates the contributions
of diagonal and off-diagonal elements of position operators.
\footnote{Even though Eqs.~(\ref{eq:csomp1}) and (\ref{eq:csomp2})
  are equivalent, the first (second) term in Eq.~(\ref{eq:csomp1})
  is not equal to the first (second) term in Eq.~(\ref{eq:csomp2}).}
(It is also manifestly invariant to the reassignment of a Wannier
function to a neighboring cell.)
The validity of Eqs.~(\ref{eq:csomp1}) and (\ref{eq:csomp2}) has been
tested numerically by comparing with the evaluation of
Eq.~(\ref{eq:theta}) for the case of a tight-binding model introduced in
Ref.~\onlinecite{malashevich10}.
The evaluated expressions agreed to numerical accuracy after
extrapolation to the infinitely dense mesh.
These expressions can also be shown to be
gauge-invariant by working directly within the Wannier representation.

\subsection{Computational details}
\label{sec:electronic}

Calculations of the electronic ground state and of structural
relaxations were performed using the %
{\sc Quantum-ESPRESSO} package,\cite{QE-2009} and the
{\sc Wannier90} code\cite{mostofi08} was used for
constructing maximally localized Wannier functions. We used
radial-grid discretized HGH\cite{hartwigsen98} norm-conserving
pseudopotentials.  Calculations were performed in the noncollinear
spin framework, including spin-orbit effects as incorporated in the
pseudopotentials.  In all calculations we used the
Perdew-Wang\cite{perdew92} LDA energy functional.
The pseudopotentials used for Cr, Fe and Gd contain semi-core
states, while the ones for Al, Bi, Se and O do not.

The self-consistent calculations on \cro\ were performed on a
$4\times 4 \times 4$ Monkhorst-Pack\cite{monkhorst76} grid in
$k$ space.
Non-self-consistent calculations for the Wannier-function
construction were
performed on $k$-space grids containing the origin and ranging in
size from $6 \times 6 \times 6$ to $12 \times 12 \times 12$. The
plane-wave energy cutoff was chosen to be 150~Ry.

In the case of \bise, the self-consistent calculations were performed on
a $6 \times 6 \times 6$ grid with energy cutoff of 60~Ry,
while the non-selfconsistent calculation was done on grids between $6
\times 6 \times 6$ and $11 \times 11 \times 11$.

The position-operator matrix elements 
$\bra{\O m}r_j\ket{\R n}$ 
needed to evaluate Eq.~(\ref{eq:csomp2})
were calculated in $k$ space by inverting
the Fourier sum in Eq.~(\ref{eq:A-R}) over the non-self-consistent
$k$-point mesh, and then approximating the $k$ derivative in
Eq.~(\ref{eq:connection}) by
finite differences on that mesh, 
as detailed in Ref.~\onlinecite{xinjie06}.

\section{Results and discussion}
\label{sec:results}

\subsection{Conventional magnetoelectrics }
\label{sec:res_conv}

In this section we present the results of our first-principles
electronic-structure calculations of \t.  We begin with
conventional magnetoelectrics, i.e., materials that are
already experimentally known to have a non-zero magnetoelectric tensor.
Some of these materials do not allow all diagonal components of the
magnetoelectric tensor to be non-zero. We omit those materials
from our analysis here, since we are interested in calculating the
CSOMP part of the magnetoelectric coupling, which would vanish
in such cases.
We first present our results on \cro\ in some detail, and then
briefly discuss our results for \bfo\ and \gdo.

\subsubsection{Calculation of \t\ in \cro}

\begin{figure}
\centering\includegraphics{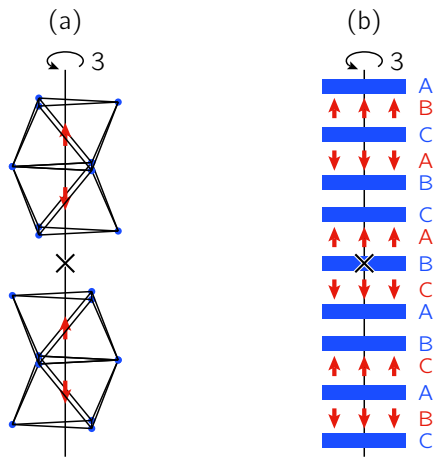}
\caption{(Color online) (a) Rhombohedral unit cell of \cro. Magnetic
  moments on Cr atoms are indicated by red arrows and oxygen
  octahedra are drawn around each Cr atom. (b) Schematic of
  hexagonal unit cell of \bise\ with imposed local Zeeman field on
  Bi atoms. Induced magnetic moments are shown
  by red arrows.  Thick blue lines indicate Se layers;
  letters (ABC) indicate stacking sequence.  In both panels,
  the vertical line indicates the 3-fold rhombohedral
  axis, and the cross designates a 2-fold rotation axis
  orthogonal to the plane of the figure
  (also a center of inversion coupled with time reversal).
  }
\label{fig:struc}
\end{figure}

We first fully relax the structure in the R$\bar{3}$c space group and
obtain the Wyckoff position to be $x=0.1575$ for Cr atoms
(4c orbit) and $x=-0.0690$ for O (6e orbit).
The length of the rhombohedral lattice vector is $a=5.3221$\,\AA\ while
the rhombohedral angle is $53.01^{\circ}$.
The Cr atoms have magnetic moments pointing along the rhombohedral
axis as illustrated in Fig.~\ref{fig:struc}(a) in an antiferromagnetic
arrangement.  The value of the magnetic moment is 2.0\,$\mu_{\rm B}$
per Cr atom and the electronic gap is 1.3\,eV, which agrees well with
previous LDA+U calculations\cite{mosey07,shi09} in
the limit where the on-site Coulomb parameter $U$ is set to zero.

Neglecting for a moment the magnetic spins on the Cr sites, the
space-group generators are a three-fold rotation, a two-fold rotation,
and an inversion symmetry as indicated in Fig.~\ref{fig:struc}(a).
Its point group is therefore $\bar{3}m$.
If we now include the spins on the Cr atoms
in the analysis, we find that the three-fold and two-fold rotations
remain, while the inversion becomes a symmetry only when
combined with time-reversal.  Therefore the magnetic point group of
\cro\ is $\bar{3}'m'$.
\footnote{Throughout the paper, the notation for magnetic point groups
  follows Ref.~\onlinecite{cracknell}.}
This magnetic point group allows $\theta$
to be different from $0$ or $\pi$, as discussed in Sec.~\ref{sec:symmetry}.

\begin{figure}
\centering\includegraphics{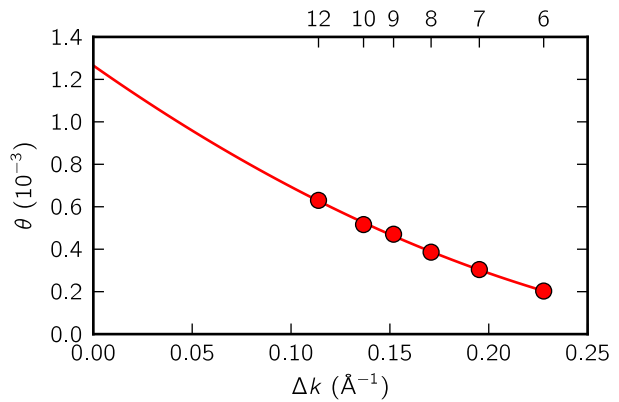}
\caption{(Color online) Calculated value of \t\ in \cro\ for varying
  densities of $k$-space grids, where $\Delta k$ is the
  nearest-neighbor distance on the grid. Top axis specifies
  the size of the corresponding uniform Monkhorst-Pack grid.
  Line indicates a quadratic
  extrapolation of $\theta$ to the infinitely dense $k$ mesh.}
\label{fig:cro}
\end{figure}

Figure~\ref{fig:cro} shows the calculated values of \t\ using
Eq.~(\ref{eq:csomp2}) for \cro\ with $k$-space meshes of various
densities. The line indicates the second-order polynomial extrapolation
to an infinitely dense mesh. The extrapolated value of \t\ is
$1.3 \times 10^{-3}$, which is a small fraction of the quantum of OMP $\theta=2
\pi$ and corresponds to $\alpha^{\rm EH}_{xx}=\alpha^{\rm
  EH}_{yy}=\alpha^{\rm EH}_{zz}=0.01$\,ps/m.
The positive sign of \t\ pertains to the pattern of Cr magnetic
moments shown in Fig.~\ref{fig:struc}(a); reversal of all magnetic
moments would flip the sign of \t.

In order to compare this value of the magnetoelectric coupling with
experimental values and other theoretical calculations, we
somewhat arbitrarily define
\beq
\alpha^{\rm eff} = \frac{ | \alpha_{xx} | + | \alpha_{yy}| + |
  \alpha_{zz}| }{3}.
\eeq
The value of $\alpha^{\rm eff}$ obtained from the results of
Ref.~\onlinecite{delaney09} is $0.23$\,ps/m
for the purely electronic part of the spin-mediated component.
Therefore, our calculated CSOMP contribution in \cro\ amounts to only
4\% of this electronic spin component.
The ionic component of the spin response calculated by the same authors
results in $\alpha^{\rm eff}=0.74$\,ps/m, while the
one calculated in Ref.~\onlinecite{iniguez08} is about 2.6 times smaller,
$0.29$\,ps/m.
(In both of these calculations,
$\alpha_{zz}$ is zero.)
Finally, experimental measurements of the magnetoelectric tensor in
\cro\ at 4.2~K
vary between $\alpha^{\rm eff}=0.55$\,ps/m and $1.17$\,ps/m
(see Refs.~\onlinecite{wiegelmann94} and
\onlinecite{kita79} respectively).

\begin{figure}
\centering\includegraphics{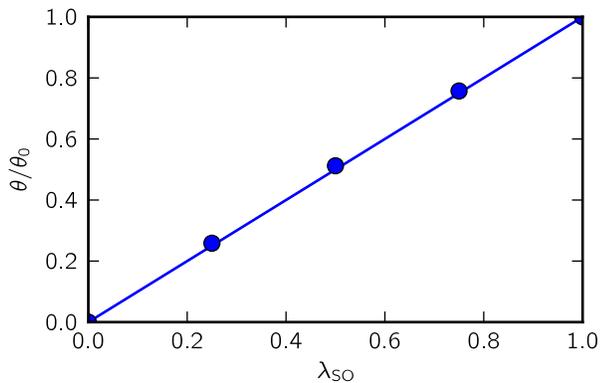}
\caption{(Color online) Calculated \t\ in \cro\ as a function of
  spin-orbit coupling strength, scaled such that $\lambda_{\rm SO}=1$
  corresponds to the full spin-orbit coupling strength and
  $\theta_0=\theta( \lambda_{\rm SO}=1)$.}
\label{fig:so}
\end{figure}

Clearly, our computed CSOMP contribution for \cro\ is negligible,
being two 
orders of magnitude smaller than the dominant
lattice-mediated spin contribution. This is probably not
surprising, since the spin-orbit coupling is relatively weak in
this material.  Given that it is weak, we can guess that that
magnitude of the CSOMP should be linear in the strength of
the spin-orbit interaction in \cro.
Our calculations allow us to check this by varying the spin-orbit
interaction strength $\lambda_{\rm SO}$ between $0$ (no spin
orbit) and $1$ (full spin-orbit interaction). As shown in
Fig.~\ref{fig:so}, if we calculate \t\ for various intermediate
values of $\lambda_{\rm SO}$, we see that the CSOMP does indeed
depend roughly linearly on $\lambda_{\rm SO}$.

\subsubsection{Other conventional magnetoelectrics}

We have also carried out calculations of \t\ in \bfo\ and \gdo, but with
a smaller number of  
$k$-point grids than in the case of \cro. Therefore, our results are less 
accurate, but should still give a correct order-of-magnitude
estimate of \t.

For \bfo\ we perform the calculation in the 10-atom
antiferromagnetic unit cell (the long-wavelength spin spiral
was suppressed).  We obtain an electronic band gap of 0.95\,eV with
magnetic moments of 3.5\,$\mu_{\rm B}$ on each Fe atom, and with a
net magnetization of 0.1\,$\mu_{\rm B}$ per
10-atom primitive unit cell
due to the canting of the Fe magnetic moments.
Extrapolating \t\ to an infinitely dense mesh using just
$6 \times 6 \times 6$ and $8 \times 8 \times 8$ $k$-point meshes, we 
obtain $\theta=0.9 \times 10^{-4}$.
In the case of \gdo\ we calculate the electronic band gap to be 5.0\,eV and
the Gd magnetic moment to be 6.7\,$\mu_{\rm B}$.  We obtain a
value of $\theta=1.1 \times 10^{-4}$ after extrapolating calculations
using $4 \times 4 \times 4$ and $6 \times 6 \times 6$ $k$-space meshes.
Thus, it is clear that the CSOMP is very small in both materials.

\subsection{Strong \z2 topological insulators}
\label{sec:res_z2}

We now investigate the CSOMP in the case of \bise, which is
known experimentally\cite{xia09} and theoretically\cite{zhang09}
to belong to the class of strong \z2\ topological insulators.
In the absence of broken $T$ symmetry, such a material should
have a $\theta$ of exactly $\pi$ (modulo $2\pi$).  We first confirm
this numerically.  Then, in Sec.~\ref{eq:afm}, we also study
what happens when $T$ is broken artificially by inducing
antiferromagnetic order on the Bi atoms
and tracking the resulting variation of \t.

\bise\ is known to belong to space group R$\bar{3}$m, with Bi at
a 2c site and Se at the high-symmetry 1a site as well as at a
2c site. In our calculations we find that the Wyckoff parameters for
Bi and Se are $x=0.4013$ and $0.2085$ respectively.
We also find the length of the rhombohedral lattice vector to be
$a=9.5677$\,\AA\ and the rhombohedral angle to be only
$24.77^{\circ}$. The electronic gap is calculated to be 0.4\,eV.

The generators of the R$\bar{3}$m space group are again three-fold
and two-fold rotations and inversion (point group $\bar{3}m$).
Since the system is nonmagnetic, the magnetic space group also
contains the $T$ symmetry operator, and its magnetic point group is
$\bar{3}m1'$. According to the analysis given in Sec.~\ref{sec:symmetry},
it is clear that \t\ must therefore be zero or $\pi$ (modulo $2\pi$).

Since we know that \bise\ is a strong \z2\ topological insulator, we
expect that \t\ should be equal to $\pi$ (modulo $2\pi$).  However,
special care needs to be taken in order to evaluate \t\ in such a
case, because the choice of a smooth gauge becomes problematic.
Specifically, it is known that the \z2\ topology presents an
obstruction to the construction of a Wannier representation (or
equivalently, a smooth gauge in $k$ space) that respects $T$
symmetry.\cite{Fu-PRB06,Roy-PRB09-a} Therefore, during the maximal
localization procedure, one needs to choose trial Wannier functions
that do {\it not} take the form of Kramers pairs, thereby explicitly
breaking the $T$ symmetry.\cite{soluyanov10} (It is important to note
that this choice of Wannier functions does not bias our calculation
towards having $\theta=\pi$, since the same starting choice of
$T$-symmetry-broken Wannier functions for a normal
$T$-symmetric insulator would result in $\theta=0$ up
to the numerical accuracy of the calculation.)

\begin{figure}
\centering\includegraphics{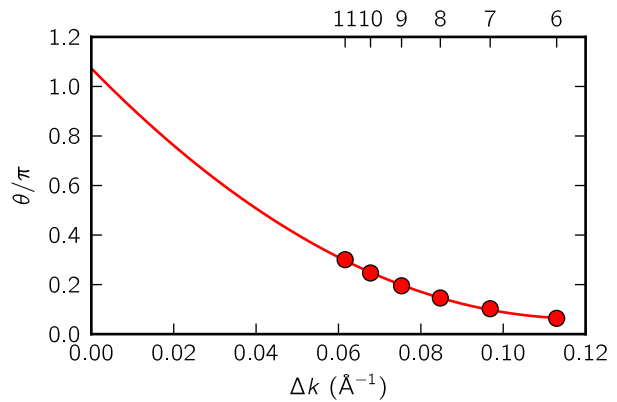}
\caption{(Color online) Calculated value of \t\ in \bise\ for varying
  densities of $k$-space grids, where $\Delta k$ is the
  nearest-neighbor distance on the grid. Top axis specifies
  the size of the corresponding uniform Monkhorst-Pack grid.
  Line indicates a quadratic
  extrapolation of $\theta$ to the infinitely dense $k$ mesh.}
\label{fig:bise}
\end{figure}

Our results for \t\ in \bise\ are given in Fig.~\ref{fig:bise} for
various densities of $k$ meshes, ranging from $6\times 6\times 6$ to
$11\times 11\times 11$.  A quadratic polynomial extrapolation to the
infinitely dense mesh limit gives $\theta=1.07 \pi$.
This is in reasonable agreement with the expected
value of $\theta=\pi$, given the uncertainties in the extrapolation.
(Of course, if we make a time-reversed choice of starting Wannier
functions, we obtain $\theta=-1.07 \pi$, which is consistent, within
the errors, with $\theta=-\pi$ and modulo $2\pi$ to $\theta=\pi$.)
Clearly the convergence with respect to mesh density
is somewhat slow, making a precise extrapolation difficult. The
reasons for this, and some possible paths to improvement, will be
discussed in Sec.~\ref{sec:summary}.

\subsection{\z2-derived nontopological insulators with broken symmetries}
\label{eq:afm}

Even though $\theta=\pi$ in \bise, a finite sample with $T$ symmetry
preserved everywhere, including at the surfaces, will not
exhibit any magnetoelectric coupling.  From the point of view of
the discussion in Sec.~\ref{sec:z2}, this happens because of
an exact cancellation between
$\theta=\pm\pi$ contributions coming
from the bulk ($\alpha$) and metallic surface ($\Delta$) terms
in Eq.~(\ref{eq:surf_cond}).  However, if one breaks the $T$ symmetry
in the bulk (and possibly some other bulk symmetries, as detailed in
Sec.~\ref{sec:symmetry}), the CSOMP term can become allowed.

The magnetic space group of \bise\ contains both $T$ and spatial
inversion symmetries. The presence of either by itself is enough to
insure that
$\theta=0$ or $\pi$ (modulo $2\pi)$. 
Now let us consider turning on, ``by hand,'' a local Zeeman field on
each Bi atom in the staggered arrangement shown in
Fig.~\ref{fig:struc}(b), i.e., with fields oriented parallel to the
rhombohedral axis and alternating in sign.  The induced magnetic
moments along the three-fold axis preserve both three-fold and
two-fold rotation symmetries; both inversion and $T$ symmetries are
broken, but $T$ taken together with inversion is still a symmetry.
The resulting magnetic point group of the system is again
$\bar{3}'m'$, as it was for \cro, and it does allow for a CSOMP
(the same magnetic arrangement has also been
discussed in Ref.~\onlinecite{Li10} 
in a different context).

\begin{figure}
\centering\includegraphics{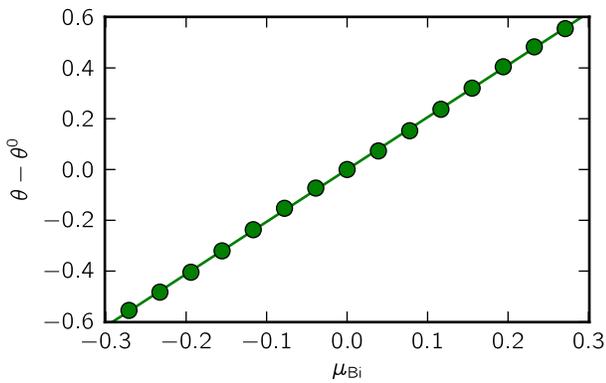}
\caption{Calculated value of \t\ (vertical axis) and induced magnetic
  moment on the Bi atom (horizontal axis) for \bise\ with artificially
  applied staggered Zeeman field on Bi atoms, as described in the
  text. $\theta^0$ is the value of CSOMP when magnetic field is not
  present.}
\label{fig:bise_afm}
\end{figure}

In the density functional calculation one can easily apply a local
Zeeman field 
to individual atoms in an arbitrary direction.  
\footnote{This is done by adding to the Kohn-Sham energy functional an
  energy penalty term of the form $\lambda \sum_i ({\bm \mu}_i -
  \bar{\bm \mu}_i)^2$, where ${\bm \mu}_i$ is the actual value of the
  magnetic moment of the $i$-th atom in the unit cell while $\bar{\bm
    \mu}_i$ and $\lambda$ are adjustable parameters.  The moments
  ${\bm \mu}_i$ are calculated by integrating the spin density within
  atom-centered spheres.}
Using this method, we have
calculated the CSOMP in \bise\ with the pattern of local fields
described previously and illustrated in Fig.~\ref{fig:struc}(b).
Fig.~\ref{fig:bise_afm} presents the calculated
values of \t\ as a function of induced magnetic moment on Bi,
where a positive $\mu_{\rm Bi}$ corresponds to the pattern of magnetic
moments indicated in Fig.~\ref{fig:struc}(b).
(Actually this was done by applying the full extrapolation procedure
of Fig.~\ref{fig:bise} for one case, $\mu_{\rm Bi}=0.16\,\mu_{\rm B}$, and
using this to scale the results calculated on the $10\times 10\times 10$
grid at other $\mu_{\rm Bi}$.)
The dependence of the change in CSOMP on the magnetic moment is linear over
a wide range.  One can see that for a relatively moderate
magnetic moment of $\pm$0.27\,$\mu_{\rm B}$, the value of \t\ is
changed from $\pi$ to $\pi \pm$0.55.
(For much higher local magnetic fields,
\bise\ becomes metallic and the CSOMP becomes ill-defined.)

These results indicate that it is possible, at least in principle,
for a magnetic material to have a large but unquantized value
of \t, thereby providing an incentive for future searches for
materials in which such a state arises spontaneously, without
the need to apply perturbations by hand as done here.

\section{Summary and outlook}
\label{sec:summary}

In this manuscript, we have presented a first-principles
method 
for calculating
the Chern-Simons orbital magnetoelectric coupling in the framework of
density-functional theory. 
We have also carried out
calculations of this coupling for a few well-known magnetoelectric
materials, namely \cro, \bfo\ and \gdo. Unfortunately,
in these materials the CSOMP contribution to the
total magnetoelectric coupling is quite small.  
This is not surprising, since in most magnetoelectric materials the
coupling is expected to be dominated by the lattice-mediated response,
whereas the CSOMP is a purely electronic (frozen-ion) contribution.
Moreover, the CSOMP is part of the orbital frozen-ion response, which
is again expected to be smaller then the spin response,
except perhaps in systems with very strong
spin-orbit coupling,
as discussed in Sec.~\ref{sec:intro}.
For example, in \cro\ the CSOMP is
about 4\% of the frozen-ion spin contribution to the magnetoelectric
coupling.

On the other hand,
 we have reasons to believe that in special cases the
CSOMP contribution to the magnetoelectric coupling could be large
compared to the total magnetoelectric coupling in known
magnetoelectrics such as \cro. 
After all, as already pointed out in Sec.~\ref{sec:largetheta},
\z2\ topological insulators are predicted to display a large
magnetoelectric effect of purely orbital origin when their
surfaces are gapped in an appropriate way.  If this is so,
why shouldn't a similar effect occur in certain $T$-broken systems?

As a proof of concept for
 the existence of those special cases, we have considered
\bise\ with inversion and time-reversal symmetries
explicitly broken ``by hand.'' Here we find that with a relatively
modest induced magnetic moment on the Bi atoms, one
can still achieve quite a large change in the CSOMP.

On the computational side, there still remain several challenges.  For
example, the convergence of our calculations of the CSOMP with respect
to the $k$-point mesh density is disappointingly slow.  A direct
calculation of \t\ in \bise\ using a very dense mesh of $11 \times 11
\times 11$ $k$ points only manages to recover about 30\% of the
converged value of $\theta=\pi$, and an extrapolation procedure is
needed to brings us within 10\% of that value.  This clearly points to
the need for methodological improvements, and we now
comment briefly on some possible paths for future work. 

The slow convergence that we observe is
related in part to the way in which we evaluate the position-operator matrix
elements $\bra{\O m}r_j\ket{\R n}$.
As discussed in Ref.~\onlinecite{xinjie06}, the
$k$-space procedure we adopted (see Sec.~\ref{sec:electronic})
entails an error of ${\cal O}(\Delta k^2)$.  Preliminary tests on a
tight-binding model suggest that an exponentially fast convergence of
\t\ can be achieved by an alternative procedure, in which the WFs are
first constructed on a real-space grid over a supercell
(whose size scales with the $k$-mesh
density), and the position matrix elements are then evaluated
directly on that grid, as in Ref.~\onlinecite{stengel06}.  It may also
be possible to improve the $k$-space calculation by using higher-order
finite-difference formulas that have a more rapid convergence with
respect to mesh density.

An alternative approach would be to develop a formula for the CSOMP
that is exactly
gauge invariant in the case of a discretized $k$-space grid. Such an
expression already exists for the case of electronic polarization,
Eq.~(\ref{eq:disc_pol}), but
we are aware of no counterpart
for the CSOMP. Even though such an approach
would not necessarily provide much faster convergence with respect to
the $k$-space sampling, it would still be a significant improvement.
For example, one would not need to construct a smooth gauge in
$k$ space, which is a particular problem in the case of \z2\
insulators (or for a symmetry-broken insulator in the vicinity of a \z2\
phase). Another use of such a formula would be to calculate with relative
ease the \z2\ index of any insulator, even in the cases when other
methods\cite{fu07,moore07,roy09} cannot be applied (for example, when
inversion symmetry is not present).

Furthermore, a full calculation of the electronic contribution to the
orbital magnetoelectric response should also include the remaining two
contributions given in Eqs.~(\ref{eq:alpha-lc}) and
(\ref{eq:alpha-ic}). This calculation would also require a knowledge
of the first derivatives of the electronic wavefunctions with respect
to electric field.  While these derivatives are available as part of
the linear-response capabilities of the {\sc Quantum-ESPRESSO}
package,\cite{QE-2009} some care is needed to arrive at a robust
implementation of Eqs.~(\ref{eq:alpha-lc}) and (\ref{eq:alpha-ic}), as will
be reported in a future communication.

Finally, recall that our calculations have all been
carried out in the context of ordinary density-functional theory.
In cases where orbital currents play a role, it is possible that
current-density functionals\cite{vignale88,vignale04} could
give an improved description.  However, such functionals
are still in an early stage of development and testing, and we
prefer to focus first on exploring the extent to
which conventional density functionals can reproduce experimental
properties of systems in which orbital currents are present.

Overall, significant progress has been made in the ability to
calculate the magnetoelectric coupling of real materials in the
context of density-functional theory.   The methods described
in Ref.~\onlinecite{iniguez08} and \onlinecite{delaney09} allow
for the calculation of both the electronic and lattice components of
the spin (i.e., Zeeman) contribution to the magnetoelectric
coupling. In principle at least, the lattice component of
the orbital contribution could be computed using the methods
of Ref.~\onlinecite{ceresoli06}, while the remaining orbital
electronic contributions can be computed from the formulas derived
in Refs.~\onlinecite{essin10} and \onlinecite{malashevich10}
following the developments discussed here.  We thus expect that
the computation of all of the various contributions
to the magnetoelectric coupling will soon be accessible to modern
density-functional methods.

\begin{acknowledgments}
We would like to acknowledge useful discussions with J.~R. Yates
and Y.~Mokrousov. The work was supported by NSF Grants
DMR-0549198, DMR-0706493, and DMR-1005838.
\end{acknowledgments}

\bibliography{pap}

\end{document}